\documentclass{aa}
\usepackage{graphicx}
\usepackage{txfonts}  
\usepackage{psfrag}

\begin{document}

\title{\object{GRB 051028}: an intrinsically faint GRB at 
high redshift?\thanks 
{Based on observations taken with the 1.34m Tautenburg telescope in
Germany, with the 2.0m Himalayan Chandra Telescope in India and with 
the 4.2m William Herschel telescope at the Spanish Observatorio del Roque 
de los Muchachos in Canary Islands.}  
}  
  
\author{A. J.    Castro-Tirado      \inst{1}  
   \and M.       Jel\'{\i}nek       \inst{1} 
   \and S. B.    Pandey             \inst{1} 
   \and S.       McBreen            \inst{2} 
   \and J.       de Jong            \inst{3} 
   \and D. K.    Sahu               \inst{4}  
   \and P.       Ferrero            \inst{5}  
   \and J.~A.~ Caballero            \inst{6} 
   \and J.       Gorosabel          \inst{1} 
   \and D. A.    Kann               \inst{5}
   \and S.       Klose              \inst{5}
   \and A.       de Ugarte Postigo  \inst{1}
   \and G. C.    Anupama            \inst{4}
   \and C.       Gry                \inst{7} 
   \and S.~ Guziy                   \inst{1,8} 
   \and S.       Srividya           \inst{4}     
   \and L.       Valdivielso        \inst{6}
   \and S.       Vanniarajan        \inst{4} 
   \and A.       A. Henden          \inst{9}
      } 
 
\offprints{A.J. Castro-Tirado, \email{ajct@iaa.es} }  
 
\institute{Instituto de Astrof\'\i sica de Andaluc\'\i a (IAA-CSIC), P.O. Box 3.004, E-18.080 Granada, Spain. 
       \and Max-Planck-Institut
  f\"{u}r extraterrestrische Physik, 85748 Garching, Germany.
       \and Max-Planck Institut f\"ur Astronomie, Koennigstuhl 17, 
            D-69117 Heidelberg, Germany.
       \and Indian Institute of Astrophysics, 560034 Bangalore, India. 
       \and Th\"uringer Landessternwarte Tautenburg, Sternwarte 5, 
            D-07778 Tautenburg, Germany.
       \and Instituto de Astrof\'{\i}sica de Canarias, Via L\'actea s/n, 
            38205 La Laguna, Tenerife, Spain. 
       \and Laboratoire d\' \rm Astrophysique de Marseille, 13376 Marseille, 
            France.
       \and Nikolaev State University, Nikolskaya 24, 54030 Nikolaev, Ukraine. 
       \and American Association of Variable Star Observers, Cambridge, MA, 
            USA.
           }

\date{Received 22 Dec 2005 / Accepted 14 Sep 2006} 
 
\abstract
  {}
  {We present multiwavelength observations of the  
  gamma-ray burst GRB 051028 detected by HETE-2   
  in order to derive its afterglow emission parameters and to 
  determine the reason for its optical faintness when compared
  to other events.}
  {Observations were taken in the optical (2.0m Himalayan Chandra  
  Telescope, 1.34m Tautenburg, 4.2m William Herschel Telescope) 
  and in X-rays (Swift/XRT) between 2.7 hours and $\sim$ 10 days 
  after the onset of the event.}
  {The data can be interpreted by collimated emission in a jet  
  with a typical value of $p$ = 2.4 which is moving in an homogeneous 
  interstellar medium and with a cooling 
  frequency $\nu_{c}$  
  still above the X-rays at 0.5 days after the burst onset. GRB 051028 
  can be classified as a ``gray'' or ``potentially dark'' GRB.      
  On the basis of the combined optical and {\it Swift}/XRT data, 
  we conclude 
  that the reason for the 
  optical dimness is not extra absorption in the host galaxy, but rather 
  the GRB taking place at high-redshift. 
  We also notice the very striking similarity with the optical 
  lightcurve of GRB 050730, a burst with a spectroscopic redshift 
  of 3.967, although  
  GRB 051028 is $\sim$ 3 mag fainter. We suggest that the bumps 
  could be explained by multiple energy injection episodes and that the
  burst is intrinsically faint when compared to the average afterglows
  detected since 1997.  The non-detection of the host galaxy down to  
  $R$ = 25.1 is also consistent with the burst arising at high redshift,  
  compatible with the published pseudo-$z$ of 3.7 $\pm$ 1.8.}
  {} 
  

\maketitle  
  
\section{Introduction}  
  
The question whether a significant fraction of gamma ray bursts (GRBs) 
are intrinsically faint or true dark events remains unsolved 
(see Filliatre et al. 2005, Castro-Tirado et al. 2006 and references therein). 
For instance, \object{GRB 000418}  
was detected in the near-IR (Klose et al. 2000) and it is one of the reddest  
(R-K = 4) together with  \object{GRB 980329} (Reichart et al. 1999), 
\object{GRB 030115} (Levan et al. 2006) and the recent 
\object{GRB 050915A} (Bloom \& Alatalo 2005). 
In most cases, it has been suggested  
that the cause of the reddening was dust extinction in the host galaxy. 
On the  other hand, \object{GRB 021211} was found to be very dim at 24 hours, 
as a scaled-down version of \object{GRB 990123} (Pandey et al. 2003).   
 
With the launch of {\it Swift} in Nov 2004, which has the ability to 
follow-up the events detected by the GRB detector onboard (BAT) or by 
other satellites like {\it HETE-2} and {\it INTEGRAL}, it is possible 
to zoom in on this population 
of optically faint events in order to disentangle their nature. 
 
\object{GRB 051028} was one of such event. It was discovered by {\it HETE-2} 
on 28 Oct 2005, lying (90\% confidence) on a 33\arcmin $\times$ 
18\arcmin error box centred at coordinates: 
RA (J2000) = 01$^{\rm h}$48$^{\rm m}$38\fs6 
Dec (J2000) = +47\degr48\arcmin30\farcs0 (Hurley et al. 2005). 
The burst started at T$_{0}$ = 13:36:01.47 UT and a value of  T$_{90}$ 
= 16~s is derived,  
putting it in the ``long-duration'' class of GRBs.  
It had   
a fluence of 6 $\times10^{-7}$ erg cm$^{-2}$ in the 2-30 keV  
range and 6 $\times10^{-6}$ erg cm$^{-2}$ in the 30-400 keV range  
(Hurley et al. 2005). 
This event was also detected by {\it Konus}/WIND 
in the 20 keV - 2 MeV range, with a duration of  $\approx12$~s, a  
fluence of (6.78$_{-1.08}^{+0.61}$)   $  
\times10^{-6}$ erg cm$^{-2}$ in the 
20 keV - 2 MeV  range and a peak energy E$_{p}$ =  298$_{-50}^{+73}$  
   keV (Golenetskii et al. 2005). 
{\it Swift}/XRT started to observe the field $\sim$ 7.1 hours 
after the event  
and detected the X-ray afterglow 5.2\arcmin ~away from the center of  
the initial ${\it HETE-2}$ error box (Racusin et al. 2005).  
 
We report here results of multi-wavelength observations in optical and 
X-ray waveband and discuss the reasons for the apparent optical faintness  
of GRB 051028 in comparison with other bursts.

\begin{table*}  
      \begin{center}  
           \caption{Journal of optical observations of the \object{GRB 051028} field.}  
                     \begin{tabular}{@{}lcccc@{}}  
  
Date of 2005 UT & Telescope/ & Filter & Exposure Time & Magnitude\\  
(mid exposure)  & Instrument &        &    (seconds)  &          \\  
\hline  
Oct 28,  16:18   & 2.0 HCT (HFOSC)  & $R_{c}$ &     300   &  20.63$\pm$0.04 \\
Oct 28,  16:32   & 2.0 HCT (HFOSC)  & $R_{c}$ &     300   &  20.72$\pm$0.05 \\
Oct 28,  16:47   & 2.0 HCT (HFOSC)  & $R_{c}$ &     300   &  21.14$\pm$0.07 \\
Oct 28,  17:03   & 2.0 HCT (HFOSC)  & $R_{c}$ &     300   &  21.27$\pm$0.07 \\
Oct 28,  17:43   & 1.34 Taut (CCD)  & $R_{c}$ &   1\,080  &  21.23$\pm$0.13 \\
Oct 28,  17:47   & 2.0 HCT (HFOSC)  & $R_{c}$ &     300   &  21.17$\pm$0.08 \\
Oct 28,  21:42   & 4.2 WHT (PFC)    & $R$     &     300   &  21.97$\pm$0.05 \\
Oct 29,  05:47   & 4.2 WHT (PFC)    & $R$     &     120   &  22.8$\pm$0.3   \\
Oct 29,  20:15   & 4.2 WHT (PFC)    & $R$     &     720   &    $>$23.7      \\
Oct 31,  22:14   & 4.2 WHT (PFC)    & $R$     &   2\,700  &    $>$25.1      \\
\hline
Oct 28,  16:25   & 2.0 HCT (HFOSC)  & $I_{c}$ &     300   &  19.79$\pm$0.11 \\
Oct 28,  16:39   & 2.0 HCT (HFOSC)  & $I_{c}$ &     300   &  19.94$\pm$0.06 \\
Oct 28,  16:55   & 2.0 HCT (HFOSC)  & $I_{c}$ &     300   &  20.29$\pm$0.09 \\
Oct 28,  17:09   & 1.34 Taut (CCD)  & $I_{c}$ &   1\,080  &  20.5 $\pm$ 0.3 \\
Oct 28,  17:10   & 2.0 HCT (HFOSC)  & $I_{c}$ &     300   &  20.38$\pm$0.08 \\
Oct 28,  17:55   & 2.0 HCT (HFOSC)  & $I_{c}$ &     300   &  20.35$\pm$0.09 \\
Oct 28,  19:12   & 1.34 Taut (CCD)  & $I_{c}$ &   1\,800  &  20.67$\pm$0.23 \\
Oct 28,  20:33   & 1.34 Taut (CCD)  & $I_{c}$ &   3\,600  &  20.75$\pm$0.13 \\
Oct 28,  22:50   & 1.34 Taut (CCD)  & $I_{c}$ &   5\,400  &  21.16$\pm$0.16 \\
\hline
Oct 28,  18:28   & 1.34 Taut (CCD)  & $V$     &   1\,080  &  22.08$\pm$0.20 \\
\hline  
                         \label{tabla1}  
                     \end{tabular}  
      \end{center}  
\end{table*}

\section{Observations and data reduction}  
  \label{observaciones}

 \subsection{X-ray observations}    
  We availed ourselves of the public X-ray observations from 
  {\it Swift}/XRT which 
  consists of four observations starting $\sim$ 7.1, 120, 150 and 230 hours 
  after the event respectively.  The detection in the first observation is 
  significant 
  (signal-to-noise ratio S/N$\sim$13.5), but in later observations the X-ray 
  afterglow is weaker and it is detected with a signal-to-noise of 3.3, 
  2.9 and 2.7.

  The XRT data is in photon counting mode and were reduced 
  using the standard pipeline for XRT data using {\it Swift} software 
  version 2.2\footnote{http://swift.gsfc.nasa.gov
/docs/software/lheasoft/download.html}  and using the most recent 
  calibration files. The data were analysed with the XSPEC version 11.3 
  (Arnaud 1996). Source and background regions were 
  extracted using a circular aperture. Spectra were selected to have
  at least 20 counts/bin.  

\subsection{Optical observations}
  Target of Opportunity (ToO)  observations  in the optical were triggered   
  starting 2.7~hours  after the event at the 2.0~m Himalayan Chandra  
  Telescope (HCT) at Indian Astronomical Observatory (HCO). 
  10\arcmin $\times$ 10\arcmin ~frames were 
  taken in imaging mode with the Himalaya Faint Object Spectrograph 
  (HFOSC), covering only 
  the central part of the large (33\arcmin $\times$ 18\arcmin)  
  {\it HETE-2} error box. Additional observations 
  were conducted at the 1.34m Schmidt telescope in Tautenburg  
  (providing a 42\arcmin $\times$ 42\arcmin ~FOV and thus
  covering the large error box) and  
  at the 4.2 m William Herschel Telescope (WHT + Prime Focus Camera)  
  at Observatorio del Roque de los Muchachos in La Palma (Spain).  
  A mosaic of 2 images (15\arcmin $\times$ 15\arcmin ~FOV) 
  were taken in order to cover the entire {\it HETE-2} error box.  
  Subsequently,  follow-up observations were taken on the following  
  days at the 4.2~m WHT.  
  Table \ref{tabla1} displays the observing log. 
  The optical field was calibrated  using the calibration files provided by 
  Henden (2005).

\section{Results and discussion}  
  \label{resultados}

\subsection{The X-ray afterglow}  
  \label{resultados}  
   
  The X-ray data confirm the presence of a decaying X-ray source in the  
  fraction (70 \%) of the {\it HETE-2} error box co\-vered by the 
  {\it Swift}/XRT, as previously reported by Racusin et al. (2005). 
  The X-ray position is RA(J2000) = 01$^{\rm h}$48$^{\rm m}$15\fs1, 
  Dec(J2000) = +47\degr45\arcmin12\farcs9   (l$^{II}$ = 132$\degr$.72, 
  b$^{II}$ = -14$\degr$.03), with an estimated uncertainty of 3\farcs8 
  (90\% containment, Page et al 2005). 
 
  The X-ray light curve in the energy range 0.3 to 10~keV is shown 
  in Fig. 1. 
  The early X-ray light curve (2$\times{10^4}$ to 5$\times{10^4}$~s) can be
  fit by a power-law decay $F_X\propto t^{\alpha_X}$
  with exponent $\alpha_X$ = $-$1.43$\pm0.60$ with a $\chi^{2}$/d.o.f = 9.3/10.
  The data were also fit including the late time data up to 10 days 
  ($\sim$8.6$\times$10$^{5}$~s) and resulted in a exponent 
  $\alpha_X$ = $-$1.1$_{-0.2}^{+0.15}$ 
  with  $\chi^{2}$/d.o.f = 10.7/13) compatible with the power-law index
  obtained considering only the early observations.  
  The value of $\alpha_X$ is dominated by the late time data 
  and a break or flattening of the
  light curve at intervening times is possible and  cannot be excluded by the 
  observations. 
 
  A spectrum was extracted for the first observation starting at 
  7.1 hours consisting of 5 {\it Swift} orbits.   
   The X-ray spectrum was fit by an 
  absorbed power-law with photon index  
  $\Gamma$=2.3$_{-0.25}^{+0.30}$ with a 
  column density  N$_{\rm H}$ = 0.40$_{-0.25}^{+0.30}$ 
  $\times$ 10$^{22}$ cm$^{-2}$ 
  (with $\chi^{2}$/d.o.f = 9.1/9) (Fig. 2). 
  The galactic column density, N$_{\rm H, GAL}$, was estimated to
  be 1.2$\times10^{21}$cm$^{-2}$ using the weighted
  average of 6 points within 1\degr of the source location 
  \footnote{http://heasarc.gsfc.nasa.gov/cgi-bin/Tools/w3nh/w3nh.pl}
  (Dickey and Lockman 1990).  The  values  used to
  estimate  N$_{\rm H, GAL}$ range from 1.01$\times$10$^{21}$cm$^{-2}$
  to 1.33$\times$10$^{21}$cm$^{-2}$.
  The fitted spectrum is compatible  
  at 90\% confidence level with Galactic absorption 
  of 1.2$\times$10$^{21}$cm$^{-2}$ (Fig. 3). 
  A power-law index of  $\Gamma$ = 1.7 $\pm$ 0.2   
  ($\chi^{2}$/d.o.f = 12.8/10)  (i.e. a spectral X-ray index $\beta_X$ = 
  $-$0.7 $\pm$ 0.2 with $F(\nu$) $\propto$ $\nu^{\beta}$) is obtained
  if only  Galactic absorption N$_{\rm H, GAL}$ of 1.2$\times10^{21}$cm$^{-2}$ 
  is considered in agreement with Page et al. (2005).  
  Alternatively, if we assume that all of the extra absorption originates in 
  the host galaxy and freeze the N$_{\rm H, GAL}$ at 
  1.2$\times10^{21}$cm$^{-2}$ then the intrinsic absorption in the host at  
  the pseudo-z (see below) of $z$=3.7 is  N$_{\rm H, z=3.7}$ of 
  12.2$^{+18.5}_{-12.1}$$\times 10^{22}$cm$^{-2}$.

\begin{figure}  
\begin{center} 
  \resizebox{8.5cm}{!}{\includegraphics[clip]{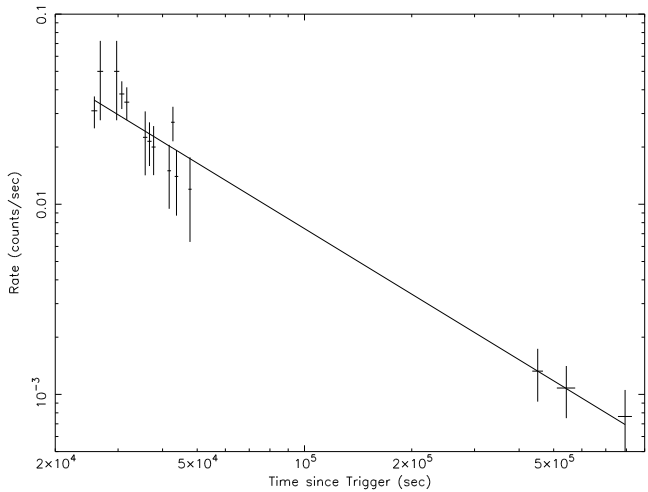}}  
 
      \caption{The X-ray lightcurve obtained by {\it Swift}/XRT  
               starting 7.1 hours after the event onset and continuing  
               up to 10 days later. The data are fit by a power-law  
               decline exponent $\alpha_{X}$ = $-$1.1$_{-0.2}^{+0.1}$.}  
    \label{x-ray lightcurve} 
\end{center}  
\end{figure}

\begin{figure}  
\begin{center} 
      \psfrag{nh}[t][r]{\LARGE nH = 10$^{22}$ cm$^{-2}$}
      \psfrag{Photon Index}{\LARGE Photon Index}
            \psfrag{(a)}{\LARGE}
       \psfrag{nH}[c]{\tiny $N_H$ $\times$ 10$^{22} $ cm$^2$} 
     \resizebox{8.5cm}{!}{\includegraphics{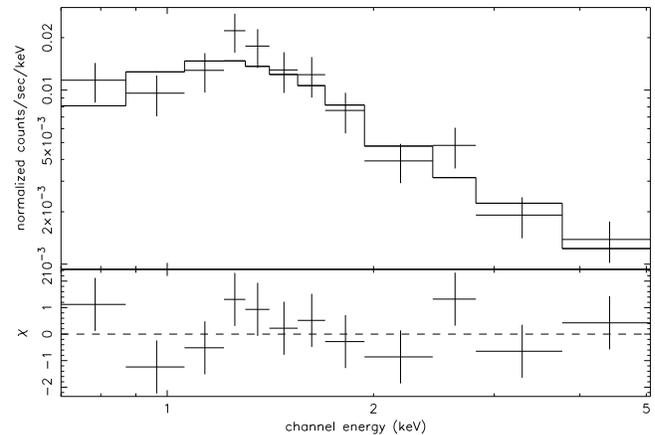}} 
      \caption{The X-ray spectrum obtained by {\it Swift}/XRT for  
               the time interval T$_{0}$ + 7.1 hours to T$_{0}$ + 13.2 hours. 
               The data can be fitted by a power-law with photon index  
               $\Gamma$ = 2.3$_{-0.25}^{+0.30}$.}  
    \label{x-ray spectrum} 
\end{center}  
\end{figure}

  \begin{figure}  
\begin{center} 
      \psfrag{NH}[l]{\LARGE  $N_{H, GAL} \sim$ 1.2$\times$10$^{21}$ cm$^{-2}$}
      \psfrag{Photon Index}{\LARGE Photon Index}
      \psfrag{nh}[c]{\Large $N_H \times$ 10$^{22}$ cm$^{-2}$} 
     \resizebox{8.5cm}{!}{\includegraphics[clip]{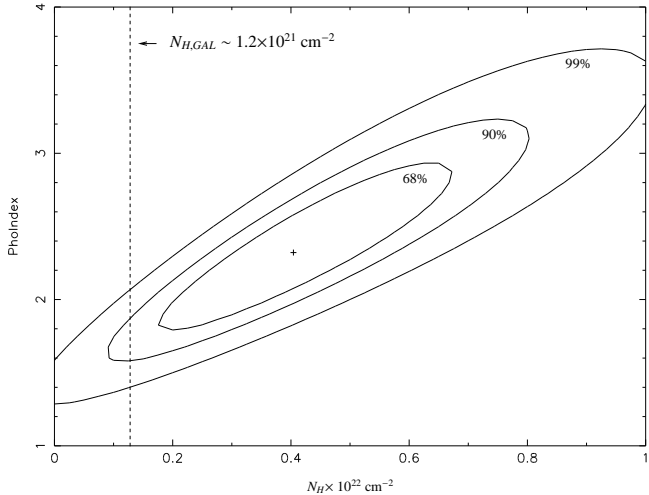}}  
        \caption {Contour plot of column 
             density versus photon index for the absorbed power law
             model shown in Figure 2.   The dashed line shows the
             estimated Galactic column density value
             1.2$\times$10$^{21}$ cm$^{-2}$ and the
             contours denote 68\%, 90\% and 99\%
             confidence levels respectively. The spectrum is compatible 
             (90\% confidence) with Galactic absorption.}
    \label{x-ray spectrum} 
\end{center}  
\end{figure} 
 
\begin{figure}  
\begin{center} 
      \resizebox{8.5cm}{!}{\includegraphics{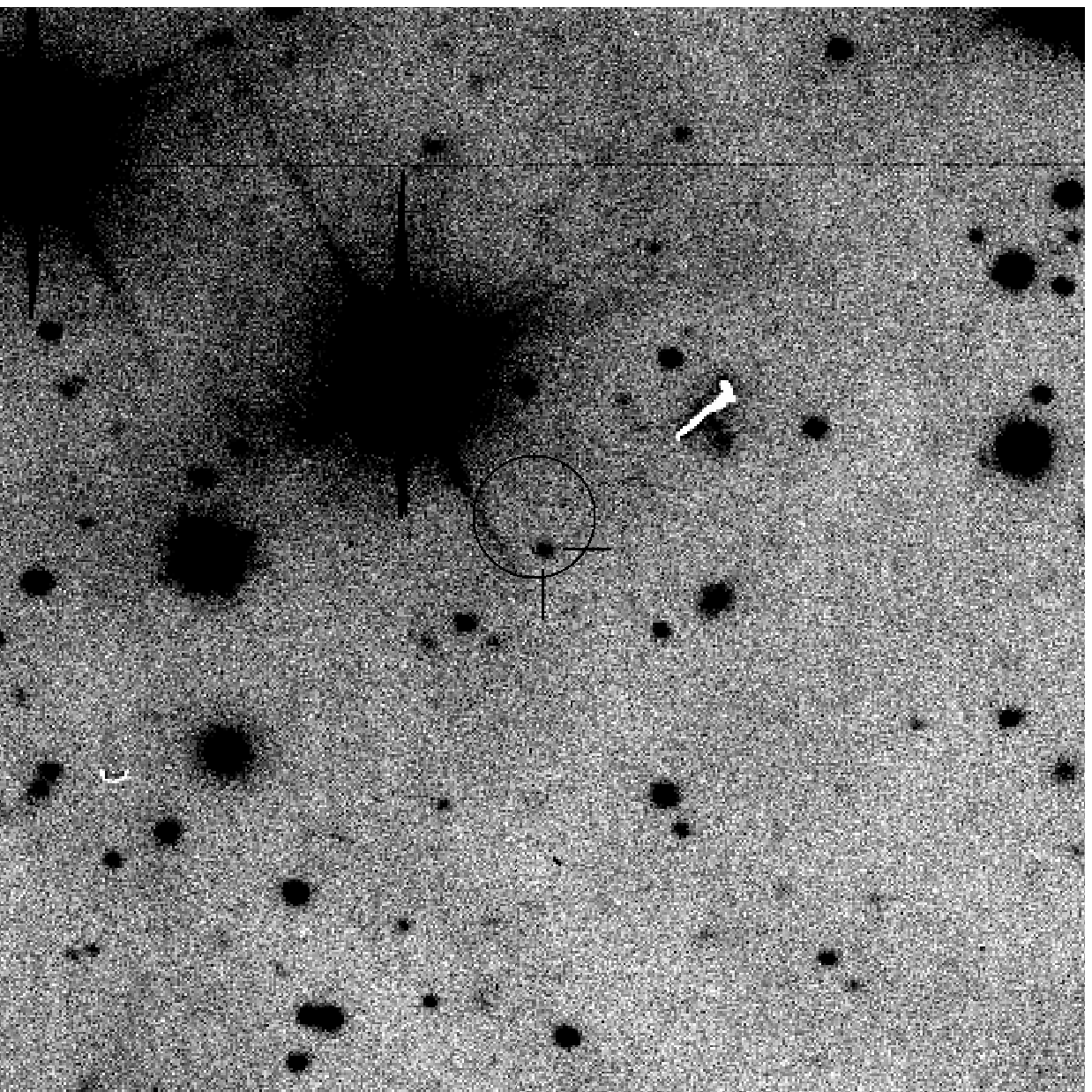}}  
      \caption{The deep $R$ band image of the \object{GRB 051028} field   
               taken at the 4.2WHT on 28 Oct 2005. The optical afterglow 
               within the 3\farcs8  {\it Swift}/XRT error box 
               (circle) is depicted. The field is 
               $2^{\prime}.5 \times 1^{\prime}.9$ with North up and East to 
               the left.}  
    \label{carta ID} 
\end{center}  
\end{figure}

    \subsection{The optical afterglow}  
             \label{sin emision}  
  
  The optical counterpart was discovered on our $R$-band images taken at 
  the 4.2m WHT telescope starting 7.5 hours after the onset of the gamma-ray 
  event. A faint $R$ = 21.9 object was detected inside the {\it Swift}/XRT 
  error circle (Jel\'{\i}nek et al. 2005, Pandey et al. 2005).  
  Astrometry against USNO-B yielded the coordinates: 
  RA(J2000) = 01$^{\rm h}$48$^{\rm m}$15\fs00, 
  Dec(J2000) = +47\degr45\arcmin09\farcs4, 
  with 0\farcs2 uncertainty (1$\sigma$, see Fig. 4).

  With E(B-V) = 0.21 in the line of sight (Schlegel et al. 1998),  
  A$_{\rm V}$ = 0.71 is derived (which translates into A$_{\rm V}$ = 0.6 
  if the correction factor proposed by Dutra et al. (2003) is taken into 
  account). A value of A$_{\rm V}$ = 0.7 is obtained using the fit from 
  Predehl and Schmitt (1995) for the Galactic H column. 
  We choose A$_{\rm V}$ $\sim$ 0.7 for the rest of this paper,
  which implies A$_{\rm R}$ = 0.53 and  A$_{\rm I}$ = 0.37. 

  From the analysis of the full $VRI$ dataset available obtained  
  at Hanle, Tautenburg and La Palma, we have obtained the optical afterglow  
  lightcurve plotted in Fig. 5.  The data between T$_{0}$ + 4 hours and  
  T$_{0}$ + 15 hours can be  
  fitted by a shallow power-law decline with decay index  
  $\alpha_{opt}$ = $-$0.9 $\pm$ 0.1.  
  The upper limits obtained at 1.5 and 3.5 day ($>$23.7 and $>$25.1 
  respectively) may suggest the existence of a break in the lightcurve  
  after $\sim$ 1 day. 
 
  The data prior to 4 hours (i.e. in the range T$_{0}$ + 2.7 hours and  
  T$_{0}$ + 4 hours)  
  show a bumpy behaviour very similar to the one seen in other events like  
  GRB 021004 (de Ugarte Postigo et al. 2005), GRB 030329 (Guziy et al. 2006  
  and references therein) and GRB 050730 (Pandey et al. 2006).  
  In fact, the similarity with GRB 050730 is very remarkable, if GRB 051028 
  is shifted up by 3 magnitudes  (Fig. 6). There is evidence for at least 
  two of such bumps taking place, superimposed on the power-law decline.
  This could be explained in the framework of multiple energy injection 
  episodes (Bj\"ornsson, Gudmundsson \& J\'ohannesson 2004). 
  GRB 050730 is an optically bright
  afterglow (see Fig. 12 of Nardini et al. 2005) whereas GRB 
  051028 seems an optically faint event if at redshift $z$ $\sim$ 3-4.
  Unfortunately there is no X-ray data available at this epoch to 
  allow a more complete modelling being carried out.

\begin{figure}  
\begin{center} 
      \resizebox{8.5cm}{!}{\includegraphics{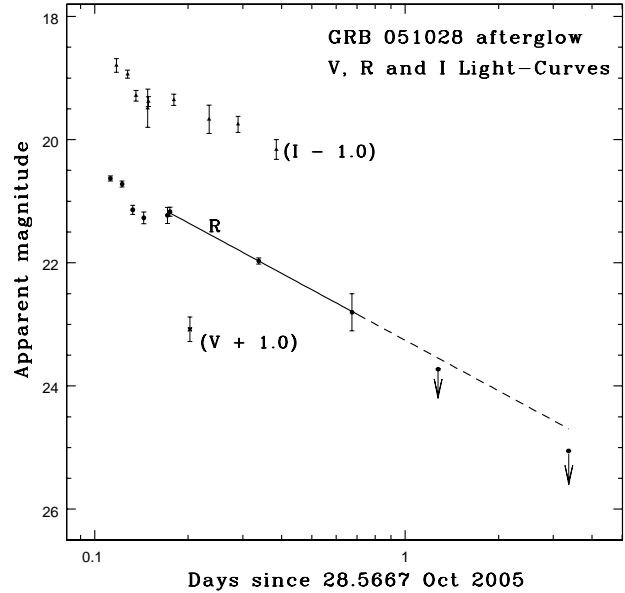}}  
      \caption{The $R$ and $I$-band lightcurves (including the $V$-band single 
               detection) obtained at Hanle (HCT), Tautenburg  
               and La Palma (WHT) starting 2.7 hours after the event onset 
               and continuing  
               up to 3.5 days later. The data after 4.0 hours are fit by a  
               power-law  
               decline exponent $\alpha_{opt}$ = $-$0.9 $\pm$ 0.1.}  
    \label{x-ray lightcurve} 
\end{center}  
\end{figure} 
 
\begin{figure}  
\begin{center} 
      \resizebox{8.5cm}{!}{\includegraphics{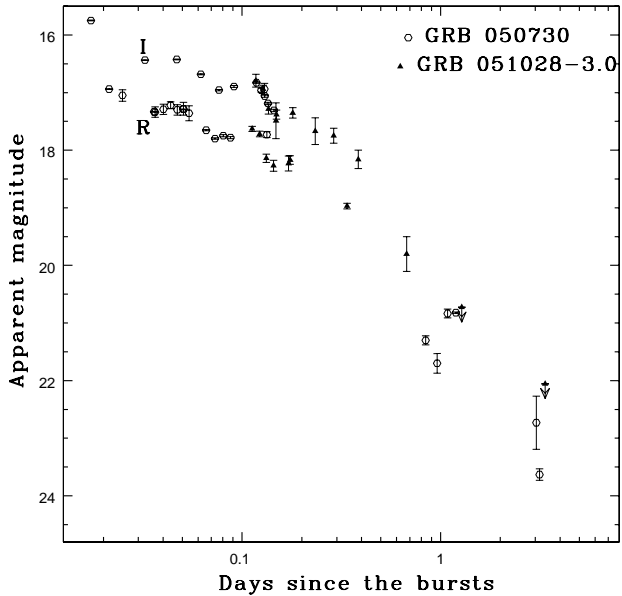}}  
      \caption{The GRB 051028 $R$ and $I$-band light curves shifted 
               by $\sim$ 3  mag in order to match the GRB 050730 
               $R$-band lightcurve (from Pandey et al. 2006). 
               No shift in the T-T$_{0}$ values has been  
               performed. These combined data strengthen the evidence of  
               a ``bumpy'' behaviour of the GRB 051028 afterglow.}  
    \label{x-ray lightcurve} 
\end{center}  
\end{figure}

    \subsection{A high redshift event}  
             \label{faintness}  
 
		  We have extrapolated the optical and X-ray fluxes of  
             the GRB 051028 afterglow to T$_{0}$ + 11 hours and 
	     derived a value of $\beta_{opt-X}$ = $-$0.55 $\pm$ 0.05. Thus  
             GRB 051028 is located in the ``gray'' or ``potentially dark''  
             GRB locus on the dark GRB diagram by Jakobsson et al. (2004). 
             How can the optical faintness of GRB 051028 be explained ?        
 
		  Although the redshift of this event could not be properly 
             measured due to its faintness at the time of the discovery, 
             we are able to constrain it on the basis of the $VRI$-band  
             data presented in this paper.  Using the magnitudes derived 
             here and correcting them for the Galactic extinction in the 
             line of sight, we determine a spectral optical index 
             $\beta_{opt}$  
	     = $-$2.1 $\pm$ 0.4. In the simplest fireball models  
	     (Sari et al. 1998), F$_{\nu}$ $\propto$ $\nu^{\beta}$ with  
             $\beta$ = $-p$/2 for $\nu$ $>$ $\nu_{c}$ and  
             $\beta$ = $-$($p$-1)/2 for $\nu$ $<$ $\nu_{c}$. 
             Thus, for a typical range of $p$ values in the range  
             1.5 $<$ $p$ $<$ 3 (Zeh, Klose \& Kann 2006), $\beta_{opt}$  
             should be in the range $-$1.5 $<$ $\beta_{opt}$ $<$ $-$0.25. 
             In fact, the GRB 051028 X-ray data before 
             T$_{0}$ + 0.5 day are well fitted by a jet model with 
             $p$ = 2.4 in the slow cooling case,  
             moving through the ISM (with $\rho$ = constant) prior to
             the jet break time 
             and with a cooling frequency $\nu_{c}$ still above the X-rays.
             A value of $\Gamma$ = 1.7 is favoured (as  $\Gamma$ = 2.3
             is giving high, unrealistic values of $p$) and thus we can 
             consider that all the absorption is Galactic in origin 
             (and ruling out dust along the line of sight in the host galaxy). 
             The X-ray data (both values of $\Gamma$) are also eventually 
             fitted for a value of $p$ = 2.1 if  $\nu_{c}$ would have already 
             crossed the X-ray band at that time (0.5 d), as it seems to be 
             derived from a sample of events studied by {\it BeppoSAX} 
             (Piro et al. 2005), but this is unlikely
             in the light of the recent {\it Swift}/XRT results for a sample
             of (presumably higher-$z$) events (Panaitescu et al. 2006). 
             In any of the above mentioned cases, the observed value of 
             $\alpha_{opt}$ can be reproduced and therefore $\beta_{opt}$ 
             should be $\sim$ $-$0.7.   What is the reason for the 
             discrepancy in the observed and expected values of  
             $\beta_{opt}$ ?   
 
		 Fig. 7 shows the derived $\beta_{opt}$  when using 
             {\it only} $VRI$ magnitudes for a sample of bursts in the range  
             3.3 $<$ $z$ $<$ 4.5 . As can be seen the derived values are in  
             the range of the one found for GRB 051028, well above the   
             $\beta_{opt}$ = 1.5 value mentioned previously. This is naturally 
             explained by the fact that at $z$ $\sim$ 3.2 and $\sim$4.0, 
             the Lyman-$\alpha$ 
             break begins affecting the $V$ and $R$ passbands respectively. 
             Therefore, one {\it natural} explanation for the $\beta_{opt}$  
             value found for GRB 051028 is that it also arose at a 
             $z$ $\approx$ 3$-$4, a value to be compared with that of 
	     GRB 050730 ($z$ = 3.967), a burst which has a suprisingly similar
	     optical afterglow lightcurve, as we have shown in Section 3.2. 
	     This $z$ $\approx$ 3$-$4 value is in fact in agreement with the 
             pseudo-$z$ = 3.7 $\pm$ 1.8 derived for this burst using the 
             recent pseudo-$z$ estimator developed by P\'elangeon et al. (2006)
             on the basis of the observed peak energy and the bolometric 
             luminosity in the 15 sec long interval containing the highest
             fluence. This would be in agreement  
             with the fact that no host galaxy is detected down to $R$ = 25.1. 
             This high-redshift is also supported by the late break time, as
             typical afterglows undergo a jet break episode before  
             T$_{0}$ + 1 day in the rest frame (Zeh, Klose \& Kann 2006).
	     In fact, the Ghirlanda et al. (2004) E$_{p}$ $-$ 
	     E$_{\gamma}$ relation is satisfied for GRB 051028 when
	     considering the pseudo-$z$ = 3.7.
 			 
             The fact that the afterglow of GRB 051028 is not 
             unusual in the {\it Swift}/XRT sample may indicate that the  
             density of the surrounding medium  
             where the progenitor has taken place should be closer to the 
             the typical value of $\approx$ 1 cm$^{-3}$ derived for several 
             long-duration GRBs. So a low density environment is not the  
             reason for its faintness at optical wavelengths. It could 
	     be that GRB 051028 could be an underluminous GRB similar to
	     {\object {GRB 980613}}, {\object {GRB 011121}} and 
	     {\object {GRB 021211}} (see Nardini et al. and references 
	     therein), in contrast to GRB 050730.

\begin{figure}[h]  
\begin{center} 
      \resizebox{8.5cm}{7.9cm}{\includegraphics{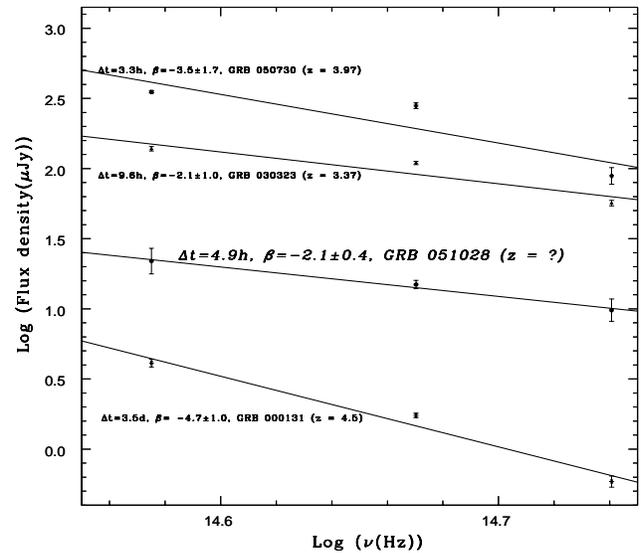}}  
      \caption{The $\beta_{opt}$ spectral indexes obtained for GRB optical 
       afterglows in the  
       redshift range 3 $<$ $z$ $<$ 4.5 based solely on the $VRI$ data. The  
       $\beta_{opt}$ value derived for GRB 051028 is in the range of those  
       derived  
       for this high-$z$ sample and therefore supports that GRB 051028 also  
       arose at a $z$ = 3-4.}  
    \label{carta ID} 
\end{center}  
\end{figure}

\section{Conclusions}  
  \label{conclusiones}  
    
  We have presented multiwavelength observations of the long duration 
  GRB 051028 detected by {\it HETE-2} between 2.7 hours and $\sim$ 10 days 
  after the event. 
  The X-ray afterglow of GRB 051028 can be compared to other 
  GRB afterglows in the sense that its flux at 11 hours is typical, i.e., one 
  can assume that the burst has occurred on a classical  n $\sim$ 1 cm$^{-3}$ 
  environment. The optical afterglow, on the other hand, is dim at a 
  similar epoch (and comparable for instance to {\object {GRB 030227}}, 
  Castro-Tirado et al. 2003). We also noticed the remarkable similarity 
  to the optical afterglow of {\object {GRB 050730}}, a burst lasting 
  $\sim$ 10 times longer with 
  comparable gamma-ray fluence\footnote{The scarcity of the available X-ray 
  data for GRB 051028 does not allow to make a straigth comparison with 
  respect to the GRB 050730 X-ray afterglow.} at $z$ = 3.967 (see Pandey et 
  al. 2006 and references therein).  
  This indicates that the 
  faintness of the optical emission is not due to a low-density environment 
  as in the case of some short GRBs, such as GRB 050509b  
  (Castro-Tirado et al. 2005). Instead,  
  we propose that GRB 051028 occurred in a faint galaxy (with $R$ $>$25.1) 
  at a high redshift consistent with the pseudo-$z$ = 3.7 $\pm$ 1.8.
   
  Thanks to the extraordinary repointing capabilities of {\it Swift}, 
  the accurate localisations for future events and the correspon\-ding  
  multiwavelength follow-up 
  will shed more light on the origin of this faint optical afterglow 
  population.

\begin{acknowledgements}  
    
  We thank the anonymous referee for useful suggestions. 
  This research has made use of data obtained through the High Energy 
  Astrophysics Science Archive Research Center On line Service, provided 
  by the NASA/Goddard Space Flight Center. Publically available 
  {\it Swift}/XRT data are also acknowledged.
  P.F., D.A.K. and S.K. thanks financial support by DFG grant Kl
  766/13-2.
  This research has also been partially supported by the Ministerio de 
  Ciencia y Tecnolog\'{\i}a under the programmes AYA2004-01515 and  
  ESP2002-04124-C03-01 (including FEDER funds).  
  
\end{acknowledgements}

\end{document}